\documentstyle[aps,twocolumn,epsf,amssymb]{revtex}
\def\ave#1{\langle #1\rangle}
\def\T{{\Bbb{T}}}
\def\R{{\Bbb{R}}}
\def\Z{{\Bbb{Z}}}

\newcommand{\half}{\textstyle{\frac{1}{2}}}
\newcommand{\sgn}{{\;\rm sgn\,}}
\begin{document}
\title{
The triangle map: a model of quantum chaos
}
\author{Giulio Casati$^{1}$, and Toma\v z Prosen$^{2}$}
\address{
$^{1}$International Center for the Study of Dynamical Systems,
Universita'degli Studi dell'Insubria,
via Lucini,3, I--22100 Como,\\
and Istituto Nazionale di Fisica della Materia and INFN, Unit\`a di
Milano, Italy,\\
$^{2}$Physics Department, Faculty of Mathematics and Physics,
University of Ljubljana, Jadranska 19, 1111 Ljubljana, Slovenia}
\date{\today}
\draft
\maketitle
\begin{abstract}
We study an area preserving parabolic map which emerges from the Poincar\' e map
of a billiard particle inside an elongated triangle. We provide numerical
evidence that the motion is ergodic and mixing. Moreover, when considered on the
cylinder, the motion appear to follow a gaussian diffusive process.
\end{abstract}
\pacs{PACS number: 05.45.-a}

The investigation of the quantum manifestations of classical dynamical
chaos has greatly improved our understanding of  the properties of quantum
motion. Even though, besides some very special cases, the non linear terms
prevent exact solution of the Schr\" odinger equation, still important useful
information can be obtained concerning statistical properties of
eigenvalues and eigenfunctions.
An important discovery has been the phenomenon of quantum dynamical
localization\cite{CCFI79} which consists in the quantum suppression of
deterministic classical diffusive behaviour. This suppression takes place
after a relaxation time scale $t_R$ which is
defined as the density $\rho$ of the operative eigenstates 
\cite{CC}, namely of those states which
enter the initial conditions and therefore determine the dynamics. For
times $t<t_R$, the quantum motion mimics the classical diffusive behaviour
and relaxation to statistical equilibrium takes place. The remarkable fact
is that quantum ``chaotic'' motion is dynamically stable as it was
illustrated in \cite{CCGS} . This means that, unlike the exponentially
unstable classical chaotic motion, in the quantum case errors in the
initial conditions propagate only linearly in time. More precisely, besides the
relaxation time scale $t_R$, a
second very important time scale exists, the so called random time scale
$t_r \sim \ln\hbar$, 
below which also the quantum motion is exponential unstable. However,
as remarked in \cite{CC} $t_r \ll t_R$ and therefore the quantum diffusion 
and relaxation process takes place in the
absence of exponentially instability. It should be noticed that, even though             
the time scale $t_r$ is very short it diverges to infinity as 
$\hbar$ goes to zero
and this ensures the transition to classical motion as required by the
correspondence principle. 

Therefore, typical quantum systems exhibit a
new type of relaxation for
which we do not have yet a physical description.
In terms of the classical ergodic hierarchy, quantum systems can be at
most mixing.  While exponential instability is sufficient for a
meaningful statistical description, it is not known whether or not it is
also necessary. Several questions remain unanswered, e.g.
there is no general relation  between the rate of exponential
instability and the decay of correlations. Moreover, as shown in
\cite{second}, quantum systems provide examples which show that linear
dynamical instability is not incompatible with exponential decay of
Poincar\' e recurrences.

In a recent paper a physical example has been found\cite{CPtri}, a billiard in
a triangle, which has zero KS entropy (the instability is only linear in time)
but which possesses the mixing property\cite{gutkin}.
This characteristics makes systems of this type, good candidates 
for the discussion of the above mentioned problems. 
In the present  paper, starting from the 
discrete bounce map for the billiard in a
triangle, we derive an area preserving, parabolic, classical map. In others
words, the map is marginally stable i.e. initially close orbits separate
linearly with time. We will show that this map is mixing, with power
law decay
of correlations and exponential decay of Poincar\' e recurrences, and
has a peculiar property: absence of periodic orbits. Moreover,
when the map is considered on the cylinder, it exhibits normal diffusion with
the corresponding Gaussian probability distribution.

Let us consider the following discontinuous 
skew-translation on the torus, with symmetric coordinates
$(x,y)\in\T^2=[-1,1)\times[-1,1)$,
\begin{eqnarray}
y_{n+1} &=& y_n + \alpha \sgn x_n + \beta \pmod{2}, \nonumber\\
x_{n+1} &=& x_n + y_{n+1} \pmod{2},
\label{eq:map2}
\end{eqnarray}  
where $\sgn x=\pm 1$ is the sign of $x$. The map                      
(\ref{eq:map2}), which we will call ``the triangle map''  
is a parabolic, piece-wise linear, one-to-one (area preserving) map,  
$\det J=1$, 
${\rm tr}\,J=2$ with 
$J:=\partial(y_{n+1},x_{n+1})/\partial(y_n,x_n)\equiv 1$. 
It is known that (continuous) irrational skew-translations 
(the above map (\ref{eq:map2}) with $\alpha=0$ and irrational 
$\beta$) are {\em uniquely ergodic} 
\cite{furstenberg} and never mixing\cite{cornfeld}, in fact they are 
equivalent to interval exchange transformations. However, the triangle map 
may have more complicated dynamics
and we show below that discontinuity  may provide a mechanism
to establish the mixing property.
Non-invertible piece-wise linear 2d parabolic maps have
been studied in Ref.\cite{zyczkowski}.

The triangle map is related to the Poincar\' e
map of the billiard inside the triangle with one angle being very small.
Indeed, let us assume that the small angle of the billiard can be written as 
$\gamma=\pi/M$ with some integer $M\gg 1$. Then the billiard dynamics may 
be {\em unfolded} by means of reflections over the two long sides of the
triangle into the 
dynamics inside a nearly-circular $2M$-sided polygon. 
Within relative accuracy of $1/M$, the approximate 
Poincar\' e map inside such polygon, 
relating two successive collisions with the short sides of 
the triangle --- the outer boundary of the polygon--- reads
\begin{eqnarray}
v_{n+1} &=& v_n + 2(u_n - [u_n] - \mu (-1)^{[u_n]}),
\label{eq:uv}\\
u_{n+1} &=& u_n - 2 v_{n+1}\nonumber
\end{eqnarray}
where $\gamma u_n$ is the polar angle and
$\gamma v_n$ is the angle of incidence of the trajectory in the 
$n-$th collision. The symbol $[x]$ is the nearest integer to $x$.
The parameter $\mu$ controls the asymmetry between the 
other two angles $\eta,\zeta$ of the triangle, namely 
$\eta,\zeta=\pi/2-\gamma (\half \pm \mu)$ and we assume that the triangle
has all angles smaller than $\pi/2$, i.e. $|\mu|\le\half$. 
As shown in \cite{CPtri}, the
system is equivalent to the mechanical problem of three elastic 
point masses on a ring (here one particle being much 
lighter than the other two).
It is interesting to note that in the scaled variables $(u,v)$ the small
parameter $\gamma$ scales out 
from the map and  the limit 
$\gamma\rightarrow 0$ simply means that the range 
of variables $u_n\in \left[0,2\pi/\gamma\right)$, 
$v_n\in \left[-\pi/(2\gamma),\pi/(2\gamma)\right)$ 
becomes the entire plane $\R^2$. The above map can be compactified 
onto a torus $\T^2$ by considering one `primitive cell' 
$(u_n\!\!\pmod{2},v_n\!\!\pmod{1})$. After transforming the coordinates as
$y_{n} = 2 (-1)^n (u_n + v_n) \pmod{2}$, 
$x_n = (-1)^n u_n - \half \pmod{2}$, 
we obtain the discontinous skew translation 
of a torus (\ref{eq:map2}) with $\alpha = 4\mu$ and $\beta=0$.

In the following we consider the general case of the triangle map with
parameters $\alpha$ and $\beta$ being two
independent irrationals. The particular case 
$\beta=0$ will be briefly discussed at the end of the paper.
We fix the parameters value 
$\alpha=(\half(\sqrt{5}-1) - e^{-1})/2$,
$\beta = (\half(\sqrt{5}-1) + e^{-1})/2$, although qualitatively identical
results
were obtained for other irrational parameter values.

\begin{figure}[htbp]
\hbox{\vbox{
\hbox{
\leavevmode
\epsfxsize=3.3in
\epsfbox{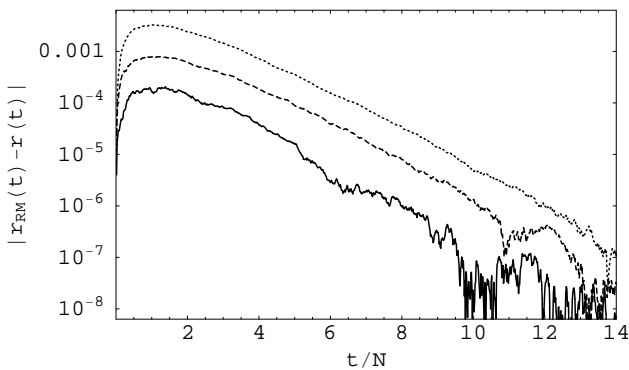}}
}}
\caption{
The deviation of the filling rate $r(t)$ from the random model in log-normal scale
for three different mesh sizes $N=10^4$ (dotted), $N=10^5$ (dashed), and
$N=10^6$ (solid curve).
}
\label{fig:1}
\end{figure}

As a first step we make a detailed and careful test of 
ergodicity of the triangle
map. To this end, following
\cite{robniketal}, we discretize the phase space $\T^2$
in a mesh of $N=N_1\times N_1$ cells and then measure the number of cells
$n(t)$ visited by  a given orbit up to discrete time $t$. 
Computing the phase space averages $\ave{.}$ by averaging over many
randomized initial conditions we compare the quantity 
$r(t) = \ave{n(t)/N}$ thus obtained with the corresponding  $r_{\rm RM}(t)$
for the {\em random model} in which  each throw onto a mesh of $N$ cells is
completely 
random. As it is known, in the latter case,  $r_{\rm RM}(t) = 1 - \exp(-t/N)$.
The result shown in fig.\ref{fig:1} provides strong evidence of
(fast) ergodicity (without any secondary time scales): namely the
exploration rate $r(t)$ of phase space for the triangle map
approaches $1$ as $t\rightarrow\infty$ and, for sufficiently fine mesh $N$, is
arbitrarily close to the 
random model $r_{\rm RM}(t)$.

\begin{figure}[htbp]
\hbox{\vbox{
\hbox{
\leavevmode
\epsfxsize=3.3in
\epsfbox{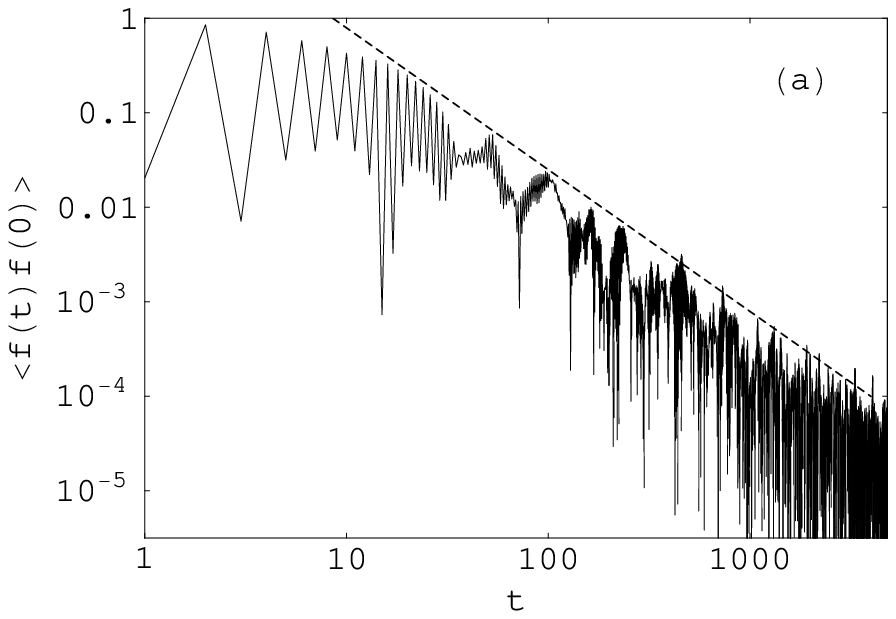}
}
\hbox{
\leavevmode
\epsfxsize=3.3in
\epsfbox{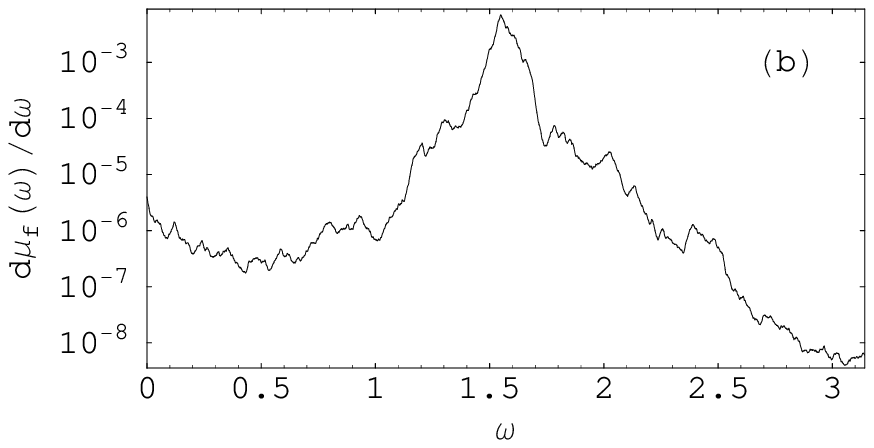}
}
}}
\caption{
The~time~auto-correlation~function~(a) 
$C(t) = \ave{\cos(\pi y_t)\cos(\pi y_0)}$  
averaged over $2\cdot 10^6$ orbits of length 16384 with
randomized initial conditions. The dashed line has slope $-3/2$.
In (b) we show the corresponding spectral density. 
Note that peak at $\omega=\pi/2$ indicates a strong component of period $4$.
}
\label{fig:2}
\end{figure}

Having established with reasonable confidence that the triangle map is
ergodic, we now turn our attention to the {\em
mixing property}. This amounts to show asymptotic decay 
of time-correlation functions of arbitrary 
$L^2$ observables.
The extensive numerical experiments we have performed, suggest that arbitrary
time-correlation
functions decay asymptotically with a power law 
$\ave{f(t)g(0)}\propto t^{-\sigma}$ with the value of the exponent $\sigma$
close to $\sigma= 3/2$.
In fig.\ref{fig:2}a we show the decay of auto-correlations of a
typical observable $f=\cos(\pi y)$.\cite{range}
The property of mixing and the nature of decay of correlations are intimately
related to the spectral properties of the {\em unitary} 
evolution (Koopman) operator over $L^2$ space of observables over $\T^2$.
The value $\sigma > 1$ we have empirically found  implies
absolutely continuous spectrum. 
Performing the inverse Fourier transform of $C(t)$:
\begin{equation}
C(t):=\ave{f(t)f(0)} = \int d\mu_f(\omega) e^{i\omega t}.
\label{eq:spc}
\end{equation}
one calculates the spectral density $d\mu_f(\omega)/d\omega$ which
should be {\em non-singular and continuous} but {\em non-smooth and
non-analytic} function, according to the (power-law, $\sigma > 1$) nature of 
decay of correlations. 
In fig.\ref{fig:2}b we show  the 
spectral density
$d\mu_f(\omega)/d\omega$ which is apparently continuous but 
not continuously differentiable function. In fact, we suggest that the
discontinuities of the derivative are {\em dense} in order to ensure the
correlation decay with the power $\sigma$ which is between $1$ and $2$.

\begin{figure}[htbp]
\hbox{\vbox{
\hbox{
\leavevmode
\epsfxsize=3.3in
\epsfbox{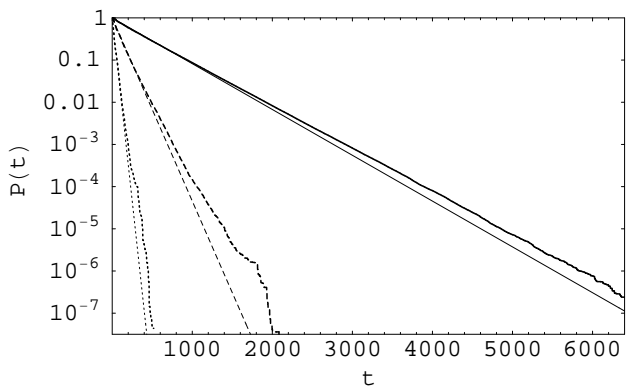}}
}}
\caption{
The Poincar\' e recurrence probabilities $P(t)$
for three different subsets $[0,0.1]\times[0,0.1]$ (solid), 
$[0,0.2]\times[0,0.2]$ (dashed), and $[0,0.4]\times[0,0.4]$ (dotted curve).
{\em Thick} curves give numerical data obtained by computing the return
probability to the subset ${\cal A}$ for a single orbit of
length 
$3\cdot 10^{11}$. {\em Thin} curves are theoretical estimates for fully
random dynamics, $P_i(t) = \exp(-\mu t)$, where $\mu$ is
the relative Lebesgue measures of the above
sets, namely $\mu=1/400,1/100,1/25$ respectively.
}
\label{fig:3}
\end{figure}

A very efficient tool for investigating the statistical properties of
dynamical
systems is the study of Poincar\' e recurrences, i.e. the probability $P(t)$
for an orbit to stay outside a specific subset ${\cal A}\subset\T^2$ for a
time longer than $t$. 
In fig.\ref{fig:3} we plot the Poincar\' e recurrence
probability $P(t)$ for the map (\ref{eq:map2}) 
and for several different subsets of the form
${\cal A}=[0,b)\times[0,b)$. The result is quite unexpected.
Indeed, for any sufficiently small set
(small $b$) the return probability appear to decay {\em exponentially}
$P(t)\propto \exp(-\lambda t)$.
Moreover,  the exponent $\lambda$ is very close to the Lebesgue measure
of the subset $\mu=|{\cal A}|$, as in the case of
the random model of completely
stochastic dynamics for which $P_{\rm RM}(t) = \exp(-\mu t)$.
Therefore the triangle map which is characterized by linear separation of orbits,
exhibits exponential decay of Poincar\' e  recurrences, typical of hyperbolic
systems. Notice that in strongly chaotic systems with positive Lyapunov
exponents, the presence of a zero measure of marginally unstable orbits
(e.g. bouncing balls in the Sinai billiard) leads to a power law decay 
of Poincar\' e recurrences.
The simultaneous presence in our model of a power law decay of correlations and
exponential decay of Poincar\' e recurrences is a fact for which, so far, we have
no explanations. Indeed, even if there are no general rigorous theorems, 
it has been conjecutured that correlations of dynamical 
observables have the same decay as the integrated Poincar\' e
recurrences\cite{artuso2}, namely 
$C(t) \sim P_i(t):=\int_t^\infty d\tau P(\tau)$ for asymptotically long 
times $t$. This relation is obviously violated in our model (\ref{eq:map2})
and this interesting point requires further investigations.

\begin{figure}[htbp]
\hbox{\vbox{
\hbox{
\leavevmode
\epsfxsize=3.3in
\epsfbox{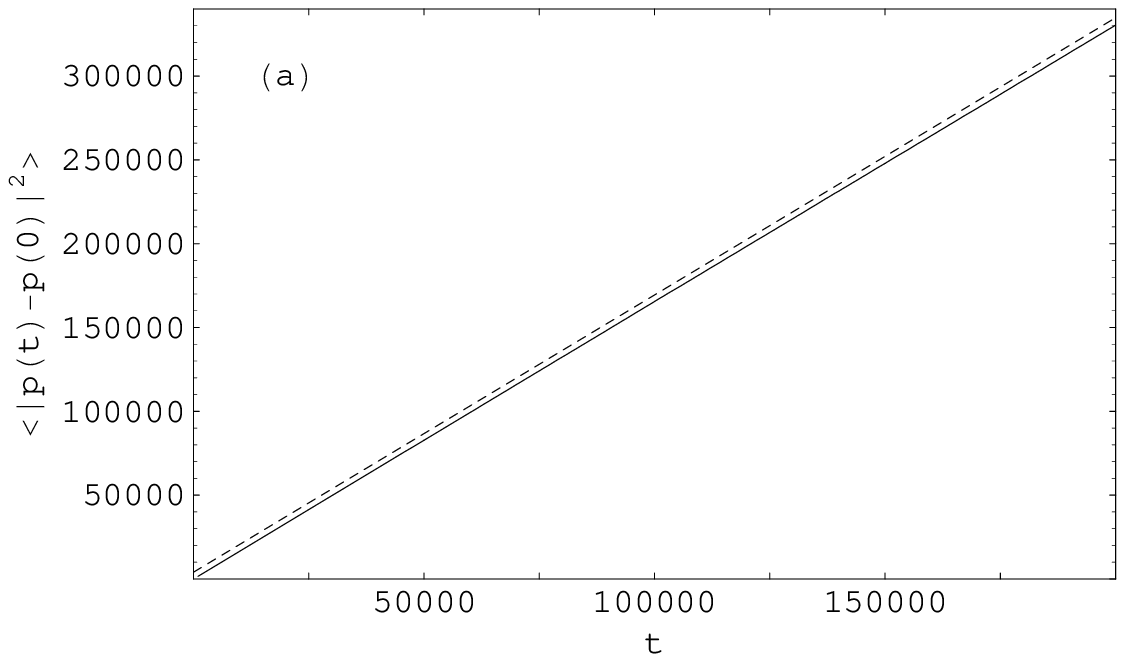}
}
\hbox{
\leavevmode
\epsfxsize=3.3in
\epsfbox{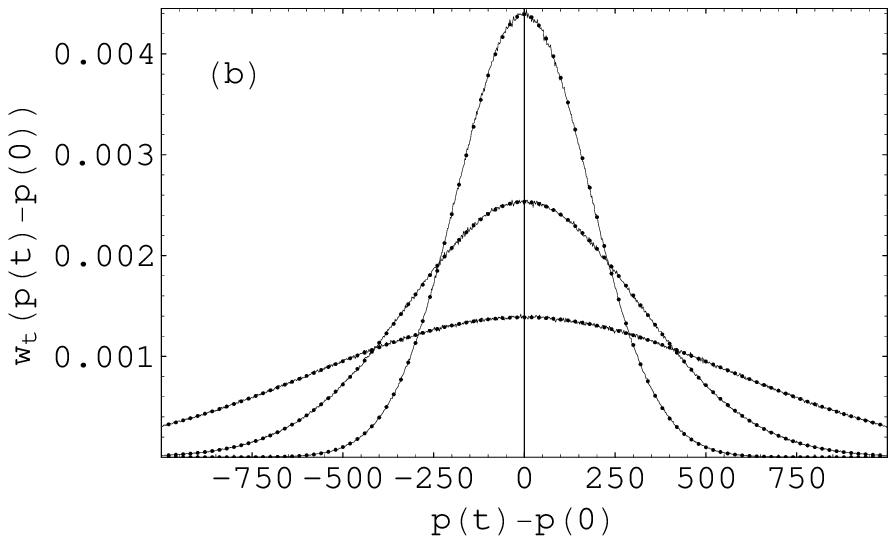}
}
}}
\caption{
The normal diffusion of the map (\ref{eq:map2}) on a cylinder. 
In (a) we show averaged squared displacement of an average 
over $10^7$ orbits of length $2\cdot 10^5$
(solid line) compared with the straight line (dashed) with 
slope $D=1.654$. In (b) we show the corresponding 
distribution of displacements at three different
times ($t=20000,60000,200000$, solid curves) which are in 
perfect agreement with the solutions of the diffusion
equation (Gaussians with variance $Dt$, dotted curves).}
\label{fig:4}
\end{figure}

Our last step is the investigation of the diffusive properties of the system. To
this end we consider the triangle map on the cylinder 
($y\in (-\infty,\infty)$). In
order to take into account the constant drift of $y_n$ 
with `velocity' $\beta$ we 
find convenient to introduce a new integer variable $p_n\in\Z$:
\begin{equation}
y_n = y_0 + \beta n + \alpha p_n,
\end{equation}
which has, by definition, vanishing initial value $p_0=0$,
and then  study the diffusive properties in the variable $p_n$.
Our numerical results shown in 
 (fig.\ref{fig:4}) provide clear numerical evidence for normal diffusive
behaviour. In particular we obtained a very accurate linear increase of 
the second moment (notice the long integration time)
\begin{equation}
\ave{(p_{n+t}-p_n)^2} = \ave{p^2_t} = D t
\end{equation}
with diffusion coefficient $D \approx 1.654$.
The almost perfect Gaussian distributions of $p_t-p_0 = p_t$ obtained at
different times (fig.\ref{fig:4}b) indicate that we are in the 
presence of a normal Gaussian process.
Note that dynamics (\ref{eq:map2}) can be rewritten in terms of a 
closed map on an integer lattice $\Z^2$ with an explicit 
`time-dependence', namely rewrite also 
the variable $x_n$ in terms of 
the integer variable $q_n\in \Z$
\begin{equation}
x_n = x_0 + y_0 n + \beta\frac{n (n+1)}{2} + \alpha q_n \pmod{2}.
\end{equation}
and the map (\ref{eq:map2}) becomes equivalent to an integer system
\begin{eqnarray}
p_{n+1} &=& p_n + (-1)^{[x_0 + y_0 n + \beta \frac{n(n+1)}{2} + 
\alpha q_n - \frac{1}{2}]},\\
q_{n+1} &=& q_n + p_{n+1},\nonumber
\end{eqnarray}
with fixed initial conditions $p_0=q_0=0$. Here, the original initial
conditions $x_0,y_0$ enter as parameters.

We note the trivial but important fact that the triangle map 
possesses {\em no periodic orbits} when the parameter $\beta$
is {\em irrational} and $\alpha$ and $\beta$ are {\em incomensurable}. 
Therefore, the general argument of Ref.\cite{CPtri} using parabolic
periodic orbits cannot be used to derive the $1/t^2$ decay of 
Poincar\' e recurrence probabilities.
It has been verified numerically that the non-existence of
periodic orbits is indeed responsible for exponential decay of
Poincar\' e recurrence probability: When we replaced irrational $\beta$
with a crude rational approximation we have obtained
a very clean crossover from initial exponential decay $\exp(-\mu t)$
to an asymptotic power law $P(t)\propto 1/t^2$ due to
existence of (long) periodic orbits.
Our map thus provides a quite  pathological example from the point
of view of semiclassical periodic orbit theory, hence we pose an
interesting question: which classical structure underpins the 
spectral fluctuations of the quantization of the triangle map 
(\ref{eq:map2})? (see also Ref.\cite{baecker} for
skew translations, $\alpha=0$).

An interesting special case of the triangle map
is $\beta=0$ which, as discussed above, 
describes  the dynamics of an elongated triangle (\ref{eq:uv}).
Here two cases should be distinguished:
i) the parameter $\alpha$ ($=4\mu$)
is {\em rational} $\alpha=2k/l$, with $k,l\in\Z$, then the dynamics is 
{\em pseudointegrable} and confined onto $l$-`valued' invariant
curves $(y_n - y_0)l \pmod{2} = 0$.
ii) the parameter $\alpha$ is {\em irrational}, then the dynamics
has been found to be {\em ergodic}. 
However ergodic properties turn out to be {\em weak} 
(see also \cite{kaplan}) and the rate of ergodicity is very slow as 
opposed to the general case $\beta\neq 0$: It has been shown that the 
number of different values of coordinate $y_n$ taken by a single orbit up to 
the discrete time $T$, $0\le n < T$, 
grows extremely slowly, as $\propto \ln T$. 
A similar property has been found for triangular billiards in which one angle 
is rationally related with $\pi$, e.g. 
right triangles \cite{artuso,CPtri}. In addition, numerically computed 
correlation functions of (\ref{eq:map2}) with
$\beta=0$, such as $\ave{\cos(\pi y_t)\cos(\pi y_0)}$, 
show perhaps a tendency to decay as power-laws but with a small 
exponent $\sigma$ around $0.1$. It is fair to say that, in this case, 
it is difficult to judge definitively, based on numerical 
experiments, on the property of mixing even though it cannot be excluded. 

In this paper we have shown that a gaussian diffusive process and mixing
behaviour can take place in a simple area preserving map without
dynamical exponential instability. One may argue that parabolic maps are non
generic and therefore irrelevant for the description of physical
systems. However, the results presented here show that a meaningfull statistical
description is possible without the strong property of exponential instability. 
Even if the model discussed here is non generic in the context of classical
systems, it can
describe the typical mechanism of quantum relaxation. Therefore it can play an
important role in understanding and describing the quantum chaotic motion
in analogy to the one played by Arnold cat map for classical systems.

T.P. acknowledges financial support by the Ministry of Science and
Technology of Slovenia. G.C. warmly thanks CIC, 
Cuernavaca (Mexico) for the kind
hospitality during the last stage of this work.


\begin{thebibliography}{99}
\bibitem{range}
We note that the time range for the plot of fig.\ref{fig:2}a has been chosen 
by comparing the numerical data with the statistical error, estimated as 
$\approx 5\cdot 10^{-5}$ at the maximal time. Due to law of large numbers the 
statistical error of $C(t)$ for fixed $t$ decays as $\sim 1/\sqrt{N T}$ 
where $N, T$ are the number, length of orbits, respectively.
Therefore, increasing integration time by factor $10$ would require
$10^{1.5}$ smaller statistical error, i.e. $10^3$ times more CPU time.
Since data of fig.\ref{fig:2} required 2 weeks of CPU time on
up-to-date workstation, results on asymptotic behaviour of $C(t)$ of 
fig.\ref{fig:2} cannot be significantly improved at present.
\end{thebibliography}
\end{document}